\documentclass[aps,superscriptaddress,amsmath,amssymb,showpacs,twocolumn,prb,reprint]{revtex4}
\usepackage{hyperref}
\usepackage{epsfig}
\usepackage{subfigure}
\usepackage{braket}
\usepackage{color}

\newcommand{\rem}[1]{}
\newcommand{\qav}[1]{\langle#1\rangle}
\newcommand{\op}[1]{\hat #1}
\newcommand{\sop}[1]{\check #1}

\newcommand{\beq}{\begin{equation}}
\newcommand{\eeq}{\end{equation}}
\newcommand{\LL}{{\check {\cal L}}}
\newcommand{\refe}[1]{\eqref{#1}}
\newcommand{\refE}[1]{Eq.~\eqref{#1}}

\newcommand{\pardiff}[2]{\frac{\partial{#1}}{\partial{#2}}}

\begin{document}

\title{Quantum current noise from a Born-Markov master equation}

\author{P. G. Kirton}
\affiliation{School of Physics and Astronomy, University of Nottingham, Nottingham NG7 2RD, United Kingdom}
\author{A. D. Armour}
\affiliation{School of Physics and Astronomy, University of Nottingham, Nottingham NG7 2RD, United Kingdom}
\author{M. Houzet}
\affiliation{SPSMS, UMR-E 9001, CEA-INAC/UJF-Grenoble 1, F-38054 Grenoble, France}
\author{F. Pistolesi}
\affiliation{Univ. Bordeaux, LOMA, UMR 5798, F-33400 Talence, France}
\affiliation{CNRS, LOMA, UMR 5798, F-33400 Talence, France}
\date{\today}

\begin{abstract}
Quantum coherent oscillations in the electric charge passing through a mesoscopic conductor can give rise to a current noise spectrum which is strongly asymmetric in frequency. The asymmetry reveals the fundamental difference between quantum and classical fluctuations in the current.  We show how the quantum current noise can be obtained starting from a Born-Markov master equation, an approach which is applicable to a wide class of systems. Our method enables us to analyse the rich behavior of the current noise associated with the double Josephson quasiparticle resonance in a superconducting single-electron transistor (SSET). The asymmetric part of the noise is found to be strongly dependent on the choice of operating point for the SSET and can be either positive or negative.  Our results are in good agreement with recent measurements.
\end{abstract}

\maketitle


{\em Introduction.}
Electrical circuits have long provided a test-bed for studying fluctuations and noise\,\cite{buttiker,nazarov:2009}.
Recent attention has focussed on the differences between quantum and classical noise far from equilibrium\,\cite{clerkRMP:2010}.
Classically, the correlation function describing the average of the product of currents measured at two times is real and
symmetric with respect to time.
In contrast, in quantum mechanics current is described by an operator which need not commute with itself at different times, leading to quantum  current correlation functions that are complex and asymmetric. The asymmetry in quantum correlation functions ultimately arises because the probabilities of a quantum system emitting or absorbing a given amount of energy are not in general the same\,\cite{nazarov:2009,clerkRMP:2010}.
The spectral density of the quantum current fluctuations, $S_I(\omega)$, is real but asymmetric in frequency and can be measured by coupling the
circuit to a mesoscopic detector and measuring the energy absorbed and emitted\,\cite{lesovik:1997,aguado:1986,clerkRMP:2010}.
Asymmetry in $S_I(\omega)$ has been observed
in superconducting tunnel junctions\,\cite{deblock:2003,billangeon:2006},
quantum point contacts\,\cite{onac:2006,gustavsson:2007}, a carbon nanotube quantum dot in the Kondo regime\,\cite{basset:2011}
and superconducting single-electron transistors (SSETs)\,\cite{billangeon:2007,billangeon:2007b,xue:2009}.

A common approach to describing transport in mesoscopic devices, especially when Coulomb blockade effects are important, is to derive a master equation for the density matrix of the system by starting from the Hamiltonian
evolution of the system coupled to leads which act as an environment.
For many devices, such as quantum dots\,\cite{gurvitz:1998,aguado:2004} or single electron transistors\,\cite{averin:1989,choi:2003,clerkNDMP:2003,blencowe:2005,clerk:2005,joyez:1994,kirton:2010}, it is then possible to use the Born-Markov approximations to describe the system using an equation of the form,
\beq
	\dot{\hat{\rho}} = \LL \hat \rho,
	\label{EqOfMotion}
\eeq
where $\hat \rho$ is the reduced density operator obtained by tracing over degrees of freedom associated with the leads and  $\LL$ is a super-operator\,\cite{cohen:1992}.

The calculation of current correlation functions is more complicated than for other quantities, such as the charge accumulated in a dot
or a spin projection, because of the difficulty of defining a current operator. Current involves keeping track of particles in the leads which do not necessarily form part of the system in a Born-Markov description.
For this reason many calculations have considered just the symmetric (classical) current noise
\,\cite{makhlin:2000,bagrets:2003,clerkNDMP:2003,novotny:2005,emary:2007},
or used equations of motion for auxiliary operators to
obtain {symmetrized} correlation functions\,\cite{choi:2003}.
Those calculations that do address the quantum noise have concentrated on effects which are important on very short time-scales where the Markov approximation fails\,\cite{engel:2004,marcos:2011,emary:2011}.

In this Rapid Communication we show how the quantum current correlation functions
can be obtained from a Born-Markov description. This approach provides a way of calculating the quantum current noise arising from coherent oscillations in the charge passing through a conductor and can be applied to a range of systems, including the SSETs probed in recent experiments\,\cite{billangeon:2007,billangeon:2007b,xue:2009}. Using our method, we provide a theoretical analysis of the quantum current noise spectrum near the double Josephson quasiparticle (DJQP) resonance.  Our results reveal a complex behavior with regions of positive and negative asymmetry in the noise, depending on the precise choice of SSET operating point, in accord with measurements\,\cite{xue:2009}.

{\em General Formalism.}
We would like to calculate
the fluctuations of the current between a given lead
and the system itself for a device whose dynamics is described by Eq.\ \eqref{EqOfMotion}.
A convenient basis for the Hilbert space is  $\{\ket{N, n} \}$,
where $\hat N \ket{N, n} = N \ket{N, n}$, with
$\hat N$ an operator counting the number of charges in the
lead, and $n$ labeling the other quantum numbers needed
to identify a given state.
In this basis the density matrix has elements:
$\rho_{\alpha}(N) = \bra{N, n} \hat{\rho} \ket{N+q , n'} $
with $\alpha=\{n,n';q\}$.
Non-diagonal elements of $\hat \rho$  with $q\neq0$
can be  non-zero
when transport is coherent, for instance
in the presence of a Josephson junction.
The equation of motion for $\rho_\alpha(N)$ reads
$
	{\dot  \rho_\alpha(N)}
	=
	\sum_{\beta,p} {\cal L}^{(p)}_{\alpha \beta} \rho_{\beta}(N+p)
$,
where we used the fact that $\LL$ cannot depend on $N$ explicitly.
This equation is solved by Fourier transforming in $N$.
Defining $\rho_\alpha(\chi)=\sum_N \text{e}^{i \chi N}\rho_\alpha(N)$ one
finds:  $\rho(\chi,t) = \text{e}^{ {\cal L}(\chi) t }\rho(\chi,0)$,
where $\rho$ is a vector with components $\rho_\alpha$ and ${\cal L}$ the matrix
${\cal L}_{\alpha \beta}=\sum_p \text{e}^{-i \chi p } {\cal L}^{(p)}_{\alpha \beta}$.

Given $\hat{\rho}$ one can obtain the (time-dependent) average of
any operator and, in particular, of $\hat N$:
\beq
	\qav{\hat{N}(t)} \equiv {\rm Tr}\{\hat N(t) \hat \rho\} =
	\left. w^{t} \frac{\partial \rho(\chi,t)}{\partial i \chi}  \right|_{\chi=0}
	\label{Neq}
	,
\eeq
where we introduced the transpose of the vector $w$, whose matrix elements
$w_\alpha$ are 1 when $\alpha=\{n,n;0\}$, that selects the diagonal elements of $\hat \rho$.
The particle current is obtained by taking a time derivative $\qav{\hat{I}(t)} \equiv -{d\qav{\hat{N}(t)}/dt}$:
\beq
	\qav{\hat{I}(t)}
	=
	-\left.
	w^{t}\left(
	{\cal L}(\chi) \pardiff{\text{e}^{{\cal L}(\chi) t}}{i \chi} 	
	+ {\cal L}'(\chi)  \text{e}^{{\cal L}(\chi) t}
	\right) \rho_0 \right|_{\chi=0}
	\,.
	\label{Ieq1}
\eeq
Here $\rho_0$ is the initial density matrix,
and a prime indicates a derivative with respect to $i\chi$.
The first term of \refE{Ieq1} vanishes since $w^t {\cal L}(0)=0$.
This relation follows from the conservation of probability
$w^{t} \rho(0, t)=1$ and the equation of motion for $\rho(\chi,t)$.
For the second term, the hypothesis that the process is
stationary implies that
$\lim_{t\rightarrow \infty} \text{e}^{{\cal L}(0)t} \rho_0=\rho_{ \rm st}$
for any initial vector $\rho_0$.
Note, however, that in general $\rho_\alpha(N,t)$
does not reach a stationary value, only
$\sum_N \rho_\alpha(N,t)$ does.
The resulting expression for the stationary current is then:
\beq
	\qav{\hat{I}}=-w^t {\cal L}'(0) \rho_{\rm st}
	\,.
\eeq

We now consider the current correlation function,
$S_I(t_1,t_2) \equiv \qav{\hat I(t_1) \hat I(t_2)} -\qav{\hat I(t_1)}\qav{\hat I(t_2)}$.
The first term can be written as follows:
\beq
	\qav{\hat I(t_1) \hat I(t_2)}
	=
	\pardiff{}{t_1}\pardiff{}{t_2}
	{\rm Tr}\{\hat N(t_1) \hat N(t_2) \hat \rho(t=0)\}
	\,.
\eeq
We now define for $t_1>t_2$
$f(t_1,t_2)=\qav{\hat N(t_1) \hat N(t_2)}$.
Since $\qav{\hat N(t_2) \hat N(t_1)}=f(t_1,t_2)^*$,
$S_I(t_1,t_2)=\partial_{t_1}\partial_{t_2}
[\theta(t_1-t_2) f(t_1,t_2)+\theta(t_2-t_1) f(t_2,t_1)^*]-\qav{\hat I(t_1)} \qav{\hat I(t_2)}$,
with $\theta(t)$ the Heaviside step function.
We can now use the quantum regression theorem\,\cite{cohen:1992} to obtain the correlator
$f(t_1,t_2)$ for $t_1>t_2$:
\beq
	f(t_1,t_2) =
	{\rm Tr} \left\{
		\check N \text{e}^{\LL (t_1-t_2)} \check N \text{e}^{ \LL t_2} \hat\rho(t=0)
	\right\}
	\label{f12}
	\,,
\eeq
where $\check N$ is the super-operator that
for any operator, $\hat A$, has the property:
$\bra{N, n} \sop{N} \op{A}\ket{M, n'} = N \bra{N, n} \op{A}\ket{M,n'} $.
Finally, we express \refE{f12} as a scalar product,
%
\beq
	f(t_1,t_2)=w^t \pardiff{}{i\chi}
	\left[
	\text{e}^{{\cal L}(\chi)(t_1-t_2)} \pardiff{}{i \chi} \text{e}^{{\cal L}(\chi) t_2}
	\right]
	\rho_0
	\,.
\eeq

Performing the time derivatives we find that
$S_I(t_1,t_2)$ only depends on $t=t_1-t_2$ for $t_2\rightarrow \infty$.
For $t\geq 0$ the correlation function reads:
\beq
	S_I(t)
	=
	\delta(t)   {\rm Re}[w^t{\cal L}''\rho_{\rm st}]
	+w^t {\cal L}'\text{e}^{{\cal L}t}{\cal L}'\rho_{\rm st}
	-\qav{\hat{I}}^2,
	\label{Stfinal}
\eeq
where the $\chi=0$ argument is omitted for brevity (the corresponding result for $t<0$ is obtained by complex conjugation).
The time Fourier transform of \refE{Stfinal}
is obtained by introducing the right and left
eigenvectors of ${\cal L}(0)$:
$ {\cal L}(0)v_\nu=\lambda_\nu v_\nu$
and $w_\nu^t {\cal L}(0)=\lambda_\nu w_\nu^t$, with
$w_\nu^t v_\mu=\delta_{\nu\mu}$.
In particular, $\lambda_0=0$ with $w_0=w$, $v_0=\rho_{\rm st}$, and ${\rm Re}\lambda_\nu < 0$ for
$\nu>0$.
We find:
\beq
	S_I(\omega) =
	{\rm Re}
	\left[
	w^t {\cal L}''  \rho_{\rm st}
	-2 \sum_{\nu\neq 0}
	\frac{w^t {\cal L}' v_\nu \, w_\nu^t {\cal L }' \rho_{ \rm st}}
	{i \omega + 		   \lambda_\nu }
	\right]\,
	\label{Somfinal}
	\,.
\eeq
%
Equation \refe{Somfinal}
is the central result of this paper: given a form for $\mathcal{L}(\chi)$, it can be used to compute the corresponding current noise spectrum.

At first sight, Eq.\ \refe{Somfinal} looks rather similar to expressions obtained by calculating probabilities for given numbers of charges, $N$, to have passed into the leads, such as in Refs.\,\onlinecite{novotny:2005} and \onlinecite{emary:2007}.
However, there is a crucial difference: the calculation in Refs.\,\onlinecite{novotny:2005} and \onlinecite{emary:2007}
assumes that there is no coherence between states with different $N$. Our choice of basis for $\rho_{\alpha}(N)$ allows us to express $\hat N$
as a simple derivative with respect to $i\chi$, leading to Eq.\ \refe{Somfinal} which does take fully into account coherence in $N$.
Including such coherences is inherently problematic for methods based on the counting statistics of
charges in the leads as discussed in Refs.\,\onlinecite{clerkNDMP:2003} and \onlinecite{belzig:2001}.

{\em Quantum Current Noise in SSETs.}
We now apply our method to the concrete example of the DJQP resonance that occurs in a SSET\,\cite{nakamura:1996,hadley:1998,thalakulam:2004,xue:2009}. A SSET consists of a superconducting island coupled by Josephson junctions to superconducting leads; a voltage applied to a gate is used to tune the island potential. When a bias voltage is applied across the device, a combination of coherent Cooper-pair oscillations and quasiparticle tunneling can give rise to a stationary current.
The DJQP resonance occurs for voltages where both resonant Cooper-pair tunneling and quasiparticle tunneling occur at both junctions\,\cite{clerk:2002,clerkNDMP:2003,clerk:2005,blencowe:2005}.
 The classical current noise for the DJQP has been calculated\,\cite{clerkNDMP:2003} (though only in certain limits).
A recent experiment that probed the asymmetry in the current noise\,\cite{xue:2009} provides motivation for us to study the full quantum problem.

\begin{figure}[t]\centering
\includegraphics[width=.35\textwidth]{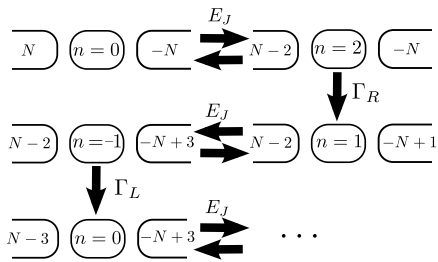}
\caption{\label{fig1}
Schematic representation of the sequence of states involved in the
DJQP cycle. The numbers give the electron
occupation of the island and of each lead.
}
\end{figure}

The Hamiltonian of the SSET is the sum of two terms: $H=H_C+H_J$.
The charging part is
\beq
     H_C=\sum_{N,n}
[E_C (n-n_g)^2 +(N+n/2) eV] \ket{N,n} \bra{N, n},
\eeq
where $V$ is the bias voltage, $E_C$ is the island charging energy and $n_g$ is the number of island charges induced by the gate voltage (the SSET is assumed to be symmetric).
The states $\{\ket{N,n}\}$  form a complete basis with
$N$ and $n$ the number of electrons in the left lead and
the island, respectively. The number of electrons in the right lead
is $-N-n$ plus a constant (which we set to zero).
The Josephson part of the Hamiltonian relevant for our problem is
\beq
H_J=-J\sum_N\left(
	 \ket{N,0}\bra{N-2,2} +\ket{N,1} \bra{N,-1} + h.c. \right),
\eeq
where $J=E_J/2$ with $E_J$ the junction Josephson energy. A DJQP resonance occurs for voltages such that $n_g = 1/2$ and $eV =2 E_C$:
the pairs of states $ \{ \ket{N,0}, \ket{N-2 ,2} \} $
and $ \{ \ket{N,1}, \ket{N,-1}\}$ are
resonant, while  quasiparticle decays are possible (provided $E_C>2\Delta/3$) between the states $\ket{N,2} \rightarrow \ket{N,1}$
through the right
junction and between $\ket{N,-1} \rightarrow \ket{N-1,0}$ through the left junction.
(Other quasiparticle decays are blocked for $E_C<2\Delta$.)
The sequence of transitions and corresponding changes in $N$ and $n$ are shown in Fig.\ \ref{fig1}.
%

%
%
%

Assuming that the normal resistance of the junctions, $R_J$, satisfies $R_J\gg h/e^2$, the dynamics is described by Eq.\
\refe{EqOfMotion} with $\LL\hat{\rho}=(\LL_{coh}+\LL_{dec})\hat{\rho}$,
where $\LL_{coh}\hat \rho=-i[\hat H,\hat \rho]$ gives the coherent evolution.
The term
$\LL_{dec}\hat \rho=\sum_{s} \Gamma_s \left[
	\ket{s'} \bra{s} \op{\rho} \ket{s} \bra{s'}
    -(1/2)\{\ket{s}\bra{s}, \op{\rho} \}
\right]$,
describes dissipative quasiparticle tunneling
as a decay process between states $\ket{s}$ and $\ket{s'}$.
The states $\ket{s} \rightarrow \ket{s'}$ concerned are
$\ket{N,-1} \rightarrow \ket{N-1,0}$
with $\Gamma_s=\Gamma_L$ and
$\ket{N,2} \rightarrow \ket{N,1}$
with $\Gamma_s=\Gamma_R$.
For given $N$, a set of 8 matrix elements of $\hat \rho$
is sufficient to describe the system\,\cite{clerk:2002,clerkNDMP:2003}.
Using the notation introduced above
we introduce the vector:
$\rho=\{\rho_{0,0;0}$, $\rho_{2,2;0}$, $\rho_{0,2;-2}$, $\rho_{2,0;2}$,
$\rho_{-1,-1;0}$, $\rho_{1,1;0}$, $\rho_{-1,1;0}$, $\rho_{1,-1;0}\}$.
Written like this, the evolution equation for the first element is
$
\dot \rho_{0,0;0}(N)
=
i J \left[\rho_{2,0;-2}(N-2)- \rho_{0,2;2}(N)\right]
+\Gamma_L \rho_{-1,-1;0}(N+1)$.
Fourier transforming with respect to $N$ gives
the first row of the matrix ${\cal L}(\chi)$:
\begin{widetext}
\newcommand{\GLm}{-i \delta_L-\frac{\Gamma_R}{2}}
\newcommand{\GLp}{i\delta_L-\frac{\Gamma_R}{2}}
\newcommand{\GRm}{-i\delta_R-\frac{\Gamma_L}{2}}
\newcommand{\GRp}{i\delta_R-\frac{\Gamma_L}{2}}
\newcommand{\eichi}[1]{\text{e}^{#1 i\chi}}
\beq
{\cal L}(\chi)=
	\left(
	\begin{array}{cccc | cccc}
	0                 & 0                 & -i J         & i J \eichi{2}     & \Gamma_L\eichi{-}    & 0 & 0 & 0 \\
	0                 & -\Gamma_R         & i J\eichi{-2} & -i J               & 0                   & 0 & 0 & 0 \\
	-i J              & i J \eichi{2}    &\GLm          & 0                  & 0                   & 0 & 0 & 0 \\
	i J \eichi{-2}     &  -i J             & 0            & \GLp               & 0                   & 0 & 0 & 0 \\
\hline
	0 & 0                  & 0 & 0 & -\Gamma_L       &  0               &  -i J              &  i J \\
	0 & \Gamma_R           & 0 & 0 & 0               &  0               &  i J               &  -i J \\
	0 & 0                  & 0 & 0 & -i J            & i J              & \GRm               & 0 \\
	0 & 0                  & 0 & 0 & i J             & -i J             & 0                  & \GRp
	\end{array}
     \right), 	
	\label{BigMatrix}
\eeq
\end{widetext}
where the detunings from the resonances for the left and right junction are defined as $\delta_L=4 E_C(n_g-1)+eV$ and $\delta_R=4 E_C n_g - eV$.
Now applying \refE{Somfinal} with $w=\{1,1,0,0,1,1,0,0\}$, numerical diagonalization
gives the frequency dependence of the current noise through the left junction, $S_L(\omega)$ for a given set of system parameters.

\begin{figure}
\centering{
\includegraphics[width=.49\linewidth]{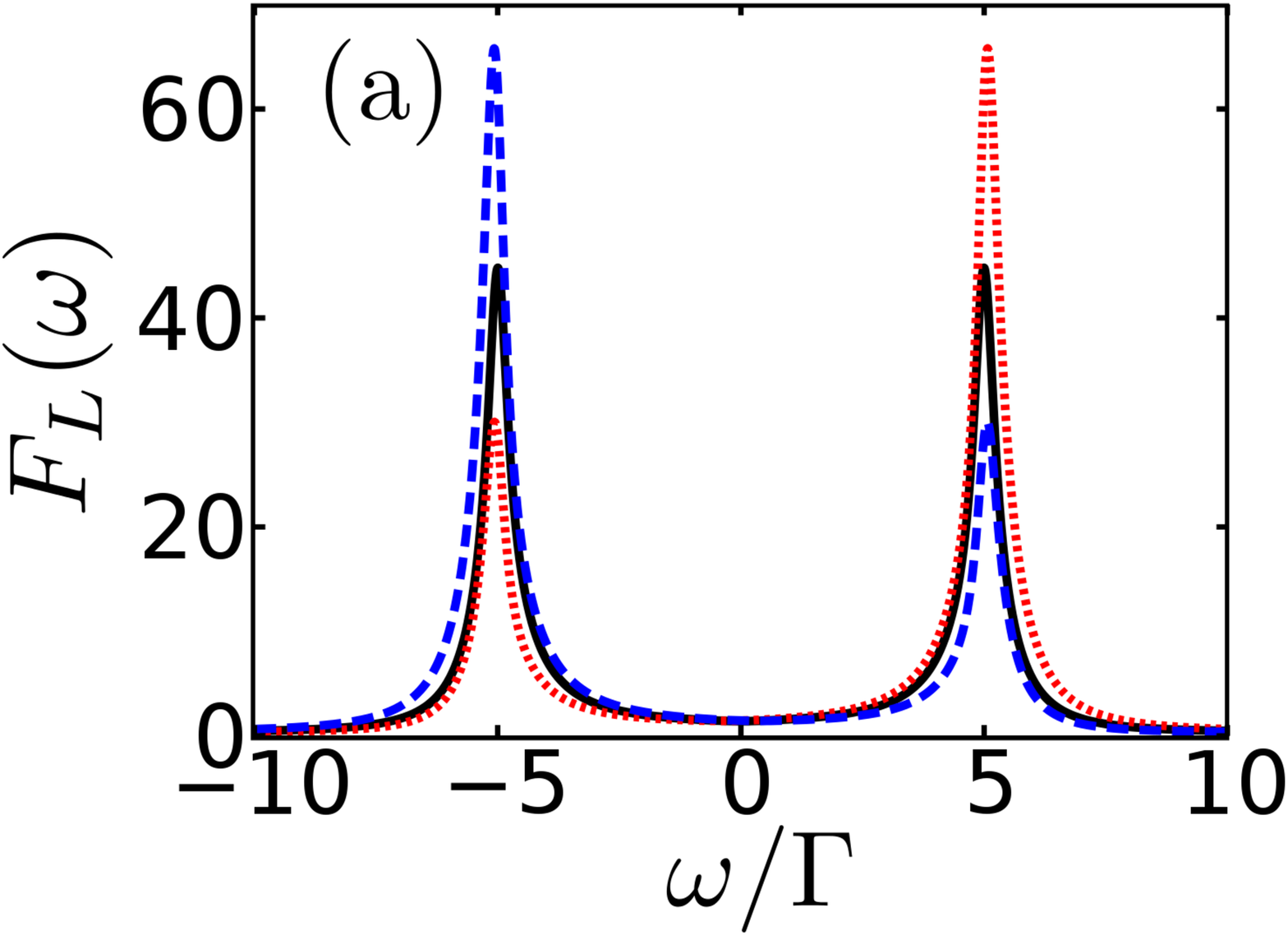}
\includegraphics[width=.49\linewidth]{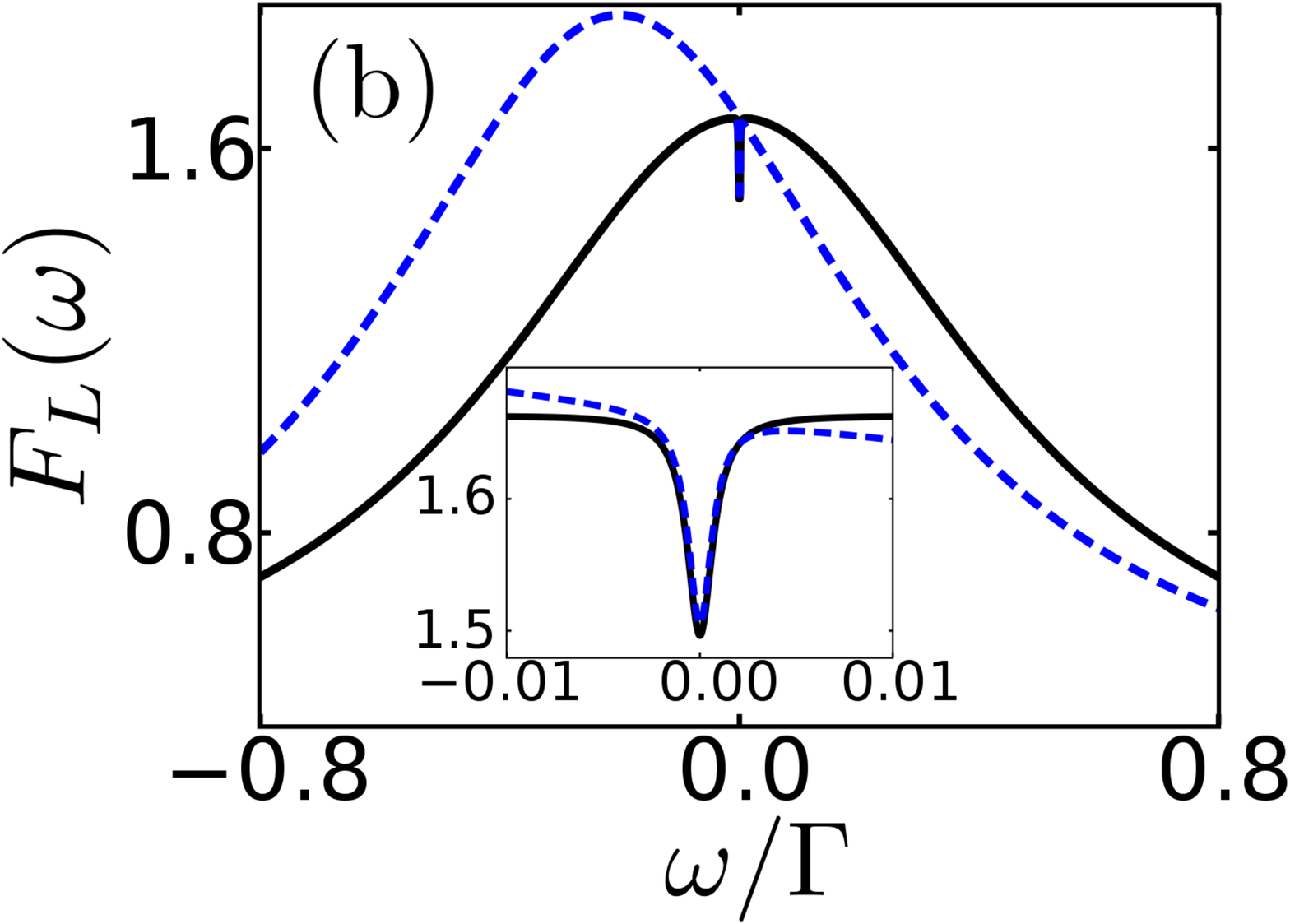}}
\caption{(Color online) Frequency dependent Fano factor, $F_L(\omega)$. (a) Weak quasiparticle tunneling regime $\Gamma/E_J=0.2$: $\delta_L=0$ (black solid line), $\delta_L=\Gamma$ (blue dashed line) and $\delta_L=-\Gamma$ (red dotted line). (b) Strong quasiparticle tunneling regime $\Gamma/E_J=50$: $\delta_L=0$ (black solid line) and  $\delta_L=0.2\Gamma$ (blue dashed line).
In all cases $\delta_R=0$ and we set $\Gamma_L=\Gamma_R=\Gamma$ neglecting the  voltage dependence of the rates.}\label{Fig2}
\end{figure}

Our quantum calculation reveals the asymmetry in the current-noise
spectrum as well as giving insights into the high frequency behavior.
Figure \ref{Fig2} shows the Fano factor for the left junction,  $F_L(\omega)=S_L(\omega)/\qav{I}$.
Defining $\Gamma$ as the quasiparticle decay rate at the center of the DJQP resonance (where $\Gamma_L=\Gamma_R$), we can
distinguish two different behaviors corresponding to weak
($\Gamma/E_J\ll 1$) and strong ($\Gamma/E_J\gg 1$) quasiparticle tunneling.
For $\Gamma/E_J\ll 1$ [see Fig.\ \ref{Fig2}(a)], coherent Cooper-pair
oscillations  lead to strong peaks in the spectrum at $\omega\simeq\pm E_J$.
At linear order in $\delta_{L(R)}$ the peaks have heights
$F_L(\pm E_J)=16 E_J(E_J\mp 2\delta_L)/(9\Gamma^2)$ and
width $\propto \Gamma$.
For  $\delta_L<0$ ($\delta_L>0$)
the positive (negative) frequency part of $F_L(\omega)$ is enhanced
since resonant oscillation in the SSET involves absorption
(emission) of energy.
The value of $\delta_R$ influences the magnitude of $F_L(\omega)$,
but not its asymmetry.
This is because $\delta_R$ affects the flow of current, but not the relative probabilities of energy absorption and emission at the left junction.
%
%
For $\Gamma/E_J\gg 1$ [Fig.\ \ref{Fig2}(b)], there are no peaks in $F_L(\omega)$, but only a dip around $\omega=0$
with $F_L=3/2$ at $|\omega|\ll E_J^2/\Gamma$ \,
and $F_L=5/3$ at $E_J^2/\Gamma\ll |\omega|\ll \Gamma$.

For arbitrary $\Gamma/E_J$ and at linear order in $\delta_{L(R)}$, $\omega$
one finds again that the asymmetry
is controlled uniquely by $\delta_L$: $F_L^{\rm asym}(\omega)=(2/3)\delta_L \omega/(\Gamma^2+4 E_J^2)$, where
we use the notation $\phi^{\rm asym}(\omega)\equiv \phi(\omega)-\phi(-\omega)$.
Note that the asymmetry has a purely quantum nature, in contrast to the symmetric part of the low frequency spectrum,
which could have been obtained using the methods of Ref. \onlinecite{clerkNDMP:2003}.

For frequencies $\omega\gg E_J,\Gamma$  we find
$F_L(\omega) \rightarrow 1/3$ independent of all system parameters
(though the Born-Markov approach breaks down eventually in the limit of very high frequencies $\omega\gg E_C$).
This sub-Poissonian noise arises because Cooper pairs contribute to the current,
but not to the high frequency noise.
The asymmetry at high frequency for small $\delta_{L(R)}$ appears with the term
$F_L^{\rm asym}(\omega)=-(4/3)(\Gamma^2+4E_J^2) \delta_L/ \omega^3$.
The noise measured by coupling a detector to one of the leads is given by a combination of the particle and displacement currents\,\cite{aguado:2004}
\beq
S_I(\omega)=[{S_L(\omega)}+{S_R(\omega)}]/{2}-{\omega^2S_Q(\omega)}/{4},
\label{fullnoise}
\eeq
where $S_Q$ is the charge noise spectrum (studied in Ref.\,\onlinecite{clerk:2005}) and we assume a gate capacitance much less than those of the junctions. The current noise at the right junction, $S_R(\omega)$, is obtained using the same technique as for $S_L(\omega)$ but using states which track the charge in the right lead.  Using a similar method,
we obtain an expression for $S_Q(\omega)$ analogous to Eq.\ \eqref{Somfinal},
\beq
S_Q(\omega)=-2\sum_{\nu\neq0}{\rm Re}\left(\frac{w^t\sop{n}v_\nu w_\nu^t\sop{n}\rho_{\rm st}}{i\omega+\lambda_\nu}\right),
\eeq
here $\sop{n}$ is a matrix corresponding to the super-operator representation of the island charge operator.

\begin{figure}
 \centering
{\includegraphics[width=.9\linewidth]{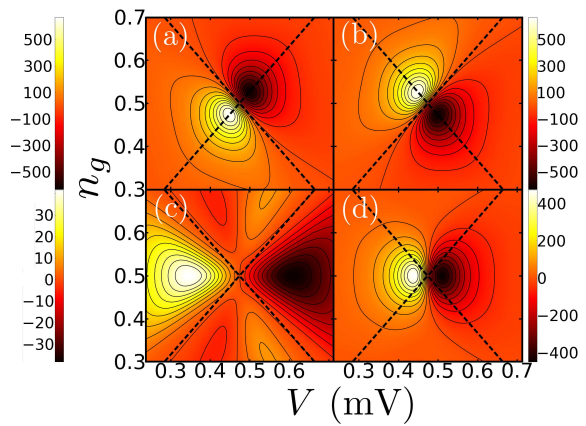}}
\caption{(Color online) Contributions to the damping (in MHz) from particle currents in (a) left, $\gamma_L$, and (b) right, $\gamma_R$, junctions; (c) contribution from the charge noise, $\gamma_Q$ and (d) total damping, $\gamma$. Dashed lines indicate the two Cooper-pair resonances which overlap at the DJQP resonance. Parameters used are: $E_C=237$\,$\mu$eV, $E_J=51$\,$\mu$eV, $\Delta=190$\,$\mu$eV, $R_J=27$\,k$\Omega$, $\omega_0/2\pi=1.04$\,GHz and $A=2.28\times10^{18}$\,C$^{-1}$\,\cite{xue:2009},
corresponding to $\Gamma/E_J\simeq 1.98$ at resonance.}\label{Fig3}
\end{figure}

In the experiments of Xue et al.\,\cite{xue:2009} an electrical resonator is used to probe asymmetry in the SSET current noise.
Within the linear-response regime\,\cite{clerkRMP:2010}, the SSET leads to damping of the resonator at the rate
 $ \gamma(\omega_0)=A^2 S^{\rm asym}_I(\omega_0)$,
where $A$ is the strength of the SSET-resonator coupling and $\omega_0$ the resonator frequency.

Using the parameters in Ref.\,\onlinecite{xue:2009} and Fermi's golden rule for calculating the quasiparticle rates\,\cite{blencowe:2005} we obtain the contributions to $\gamma$
from each of the terms in \refE{fullnoise}.
 The particle current contributions, $\gamma_{L(R)}=A^2S^{\rm asym}_{L(R)}(\omega_0)$, shown in Figs.\ \ref{Fig3}(a) and \ref{Fig3}(b), are (almost\,\footnote{There is a perfect antisymmetry when the voltage dependence of the quasiparticle rates is neglected; including this dependence, as is the case in Fig.\ \ref{Fig3}, leads to small deviations.}) antisymmetric about lines where the corresponding junction has a Cooper-pair resonance. Regions of positive (negative) damping arise when a resonance is detuned so energy is on average absorbed (emitted) from the resonator by the SSET. The charge noise contribution, $\gamma_Q=A^2\omega_0^2 S^{\rm asym}_{Q}(\omega_0)$, [Fig.\ \ref{Fig3}(c)] has a different symmetry  as both Cooper-pair resonances affect the island charge.
The overall damping, $\gamma=(\gamma_L+\gamma_R)/2-\gamma_Q/4$, shown in Fig.\ \ref{Fig3}(d), is dominated by $\gamma_{L}$ and $\gamma_{R}$; the influence of  $\gamma_Q$ is weak because the frequency scale for the SSET is set by $E_J$, which is much larger than $\omega_0$.

A simple comparison can be made with Ref.\,\onlinecite{xue:2009} by computing the maximum and minimum values of $\gamma$, which occur for $n_g=0.5$ and bias voltages below and above the center of the DJQP resonance [see Fig.\ \ref{Fig3}(d)].
We obtain  maximum and minimum damping rates with the same magnitude,  $475$\,MHz, but opposite sign, in accord with the symmetry of the problem.
Measured maximum and minimum damping rates\,\cite{xue:2009} were $\approx550$\,MHz and $\approx-35$\,MHz respectively.
 Our calculation fits with the experiment on the low bias side, though
 agreement is less good on the high bias side. The difference is probably due to the low resistance junctions ($R_J=27$\,k$\Omega$)  used\,\cite{xue:2009,thalakulam:2004} which allow higher-order processes beyond the DJQP whose contribution to the current (and hence to the damping) increases with the bias voltage.

{\em Conclusions.}
We have shown that quantum current noise in a mesoscopic conductor
can be calculated using a Born-Markov master equation description.
The theory presented allowed us to find the asymmetry of the
quantum current noise at the DJQP resonance in SSETs and
to confirm the interpretation of a recent experiment that
measured this asymmetry by detection of emission and absorption
of energy.
 The method we derived here has a wide scope of applicability. It could, for example, be applied to Cooper pairs resonances in the SSET, for which theoretical predictions have not yet been made though the quantum current noise was measured recently\,\cite{billangeon:2007}.

Funding from EPSRC (UK) under grant EP/I017828/1 (AA), and from ANR (France), under grants ANR-11-JS04-003-01 (MH) and ANR QNM No. 0404 01 (FP) is gratefully acknowledged.

\bibliographystyle{apsrev4-1}

\bibliography{biblioNEMS}
\end{document}